\documentclass[11pt]{article}
\usepackage[dvips]{geometry}
\usepackage{eurosym}
\usepackage{graphicx}
\usepackage{epstopdf}
\usepackage{amssymb,amsmath}
\usepackage{graphicx,float}
\usepackage{indentfirst}

\newcommand{\be}{\begin{equation}}
\newcommand{\ee}{\end{equation}}
\newcommand{\ba}{\begin{eqnarray}}
\newcommand{\ea}{\end{eqnarray}}

\begin{document}

\author{G. P. de Brito\thanks{gustavopazzini@gmail.com} and A. de Souza Dutra\thanks{dutra@feg.unesp.br} \\
\\
UNESP Univ Estadual Paulista - Campus de Guaratinguet\'a - DFQ\\
Av. Dr. Ariberto Pereira Cunha, 333\\
12516-410 Guaratinguet\'{a} SP Brasil}
\title{Orbit based procedure for doublets of scalar fields and the emergence
of triple kinks and other defects}
\maketitle

\begin{abstract}
In this work we offer an approach to enlarge the number of exactly solvable models with two real scalar fields in (1+1)D. We build some new two-field models, and obtain their exact orbits and exact or numerical field
configurations. It is noteworthy that a model presenting triple-kinks and double-flat-top lumps is among those new models.
\end{abstract}

\section{Introduction}
\indent The most part of the natural physical systems can be studied by using linear differential equations, with their good properties like the superposition principle. However, in the last few decades there is a growing in deal with system which are intrinsically nonlinear, specially those systems that supports topological defects. In fact, topological structures play an import role in the development in several branches of physics, from condensed matter to high energy physics and cosmology \cite{rajaraman}-\cite{vachaspati}. In condensed matter, a recent and interesting example regarding topological defects is related with the study of magnetic domain wall in a nanowire, designed for the development of magnetic memory \cite{bischof}. In high energy physics we may cite, for instance, the importance of defect structures in brane world scenarios, where we may interpret that we live in a domain-wall with 3+1 dimensions embedded in a 5-dimensional spacetime \cite{rubakov,gremm}. In cosmology, topological defects may be related with phase transitions in the early Universe, such defects may have formed as the Universe cooled and various local and global symmetries were broken \cite{vilenkin,kibble}.\\
\indent In this work we focus in models for doublets of scalar fields. It is remarkable that, whenever this models have a potential with two or more degenerate minima, one can find topological solutions connecting them. For models with a single scalar field it is usual we arrive at kink-like solutions. However, when we deal with models with two scalar fields the vacua structure may be richer, and as a consequence, other kinds of defects are possible. The so-called BNRT model \cite{bnrt}, for instance, has a vacua structure with four degenerate minima and the doublet of scalar field admits kink-like (topological) solutions for one of its components and lump-like (non-topological) to the other one. For the same model, one can find double-kinks and flat-top lumps, and also, there is a critical case where both components of the doublet are kink-like configurations \cite{Dutra1,Shifman}. As we will see, in this paper we will arrive with models that possess very interesting vacua structures, engendering kinks, double-kinks and even triple-kinks configurations.\\
\indent In fact, double and triple-kinks are particular case of the so-called multikink configurations \cite{prd2014}. The interest in deal with multikinks was, in part, motivated by the discovery of Peyrard and Kruskal \cite{peyrard} that a single kink becomes unstable when it moves in a discrete lattice at sufficiently large velocity, while multikinks remains stable. This effect is associated with the interaction between the kink and the radiation, and the resonances were already observed experimentally \cite{kinkinteraction}. Some years ago, Champney and Kivshar \cite{champney} performed an analysis on the reasons of the appearance of multikinks in dispersive nonlinear systems. Furthermore, multikinks have applications in different areas of physics. For instance: in may study of mobility hysteresis in a damped driven commensurable chain of atoms \cite{braun}. In high energy physics, double-kinks are important to explain the split-brane mechanism in braneworld scenarios \cite{hoff,splitbranes}.\\
\indent Another motivation to the study of models with two scalar fields, its related with intersection of defects and the construction of networks of defects \cite{Bazeia2,bazeiabrito}. This subject may find applications, for instance, in cosmology \cite{vilenkin}, in magnetic materials \cite{eschenfelder} and in the study pattern formation in condensed matter \cite{walgraef}. Essentially the construction of networks of defects is related with a symmetry with respect to some discrete group (\textit{e.g.} $Z_3$ or $Z_2 \times Z_2$) acting in the vacua structure. Hence, we are motivated with the possibility of construct new models for doublets of scalar fields with different vacua structures.\\
\indent Unfortunately, as a consequence of the nonlinearity, we face some troubles when we deal with these systems analytically. The framework can be simplified considerably for systems in $1+1$ dimensions, in this case we may reduce the set of second-order differential equations to a set of first-order ones, using the so-called Bogomol'nyi-Prasad-Sommerfield (BPS) procedure \cite{BPS}. However, if there is more than one scalar field in the model, those first-order differential equations are still coupled and the difficulty of solving the problem is in general great. In fact, the trial and error method historically arose due to the inherent difficulty to get general methods for solving nonlinear differential equations. Rajaraman \cite{Rajaraman2} introduced an approach of this nature for the treatment of coupled relativistic scalar field theories in $1+1$ dimensions. His procedure was model independent and can be used for search solutions in arbitrary coupled scalar fields models in $1+1$ dimensions. However, the method is convenient and profitable only in some particular, but important, cases. Some years later, Bazeia and collaborators applied the approach developed by Rajaraman to some important models \cite{Bazeia1,Bazeia3}. Some years ago, it has been noted that in the case of the coupled nonlinear first-order equations, one can obtain a differential equation relating both fields, and its solution lead to a general orbit connecting the vacua of the model \cite{Dutra1,Dutra2}. The number of exact models with two scalar fields have been enlarged a little through the so-called deformation approach \cite{Dutra5,Afonso}. Recently, it was shown that one can go further by performing a deformation of the orbit equation \cite{prieslei}. On the other hand, at least partially, one can devise the general behavior of the topological solutions of a given nonlinear model by studying its vacuum structure and its orbits \cite{meza}. In the last reference, it was noticed that the appearance of double-kink and flat-top lumps, was a consequence of the passage of the orbit in the vicinity of a vacuum, before to go to another one. As we are going to see in this work, in one of the new models introduced here, this feature will give rise to the emergence of triple-kinks and a kind of double flat-top lump. \\
\indent Despite of all advances mentioned in the last paragraph, the number of nonlinear systems with two interacting scalar fields that can be exactly solved is yet relatively small, mostly due to the difficulty in getting solutions of the orbit equations. In this work we introduce an approach in order to tackle with this kind of problem. As we are going to see, it will allow us to expand very much the number of systems with two
nonlinearly interacting scalar fields, for which one can get access to an analytical expression of the orbit equation and, as a consequence, construct solitonic configurations.\\
\indent The Lagrangian density for the case of two coupled scalar fields that we are going to work with is given by
\begin{equation}
\mathcal{L}=\frac{1}{2}(\partial _{\mu }\phi \partial ^{\mu }\phi +\partial
_{\mu }\chi \partial ^{\mu }\chi )-V(\phi ,\chi ),
\end{equation}
whose Euler-Lagrange equation for static configurations are
\begin{equation}
\frac{d^{2}\phi }{dx^{2}}=\frac{\partial V}{\partial \phi }\quad ,\quad
\frac{d^{2}\chi }{dx^{2}}=\frac{\partial V}{\partial \chi }.
\label{second_order}
\end{equation}%
An interesting consequence arises if one considers a class of potentials
that can be written in terms of a superpotential function $W(\phi ,\chi )$,
namely
\begin{equation}
V(\phi ,\chi )=\frac{1}{2}(W_{\phi }^{2}+W_{\chi }^{2}),
\end{equation}%
in such case we are able to get a first order formalism to solve the
problem, in fact, it is easy to verify that the following equations share
the same solutions of (\ref{second_order})
\begin{equation}
\frac{d\phi }{dx}=W_{\phi }\quad ,\quad \frac{d\chi }{dx}=W_{\chi }.
\label{firstorder}
\end{equation}%
In general, the above equations are coupled through nonlinear terms. So, the
usual methods of linear algebra are not useful in this case. In order to
turn the above system decoupled, we note that it is possible to combine both
equations to get the so called orbit equation
\begin{equation}
\frac{d\phi }{d\chi }=\frac{W_{\phi }}{W_{\chi }},  \label{orbit}
\end{equation}%
once we solve this equation, we are going to be able to write $\phi (\chi )$
or $\chi (\phi )$ and, then, eliminate one of the fields on (\ref{firstorder}%
).

\section{The method}

In this section we are going to present a method to solve orbit equations.
For this we consider an implicit solution of orbit equation given by $F(\phi
,\chi )=c$, where $c$ is a constant. Now, let us differentiate $F(\phi ,\chi
)$ to obtain
\begin{equation}
dF(\phi ,\chi )=\frac{\partial F}{\partial \phi }d\phi +\frac{\partial F}{%
\partial \chi }d\chi =0,  \label{dF}
\end{equation}%
we will also consider the orbit equation rewritten as $W_{\chi }d\phi
-W_{\phi }d\chi =0$. Moreover, one can multiply this equation by an
integrating factor $H(\phi )$ in order to get
\begin{equation}
H(\phi )W_{\chi }d\phi -H(\phi )W_{\phi }d\chi =0.  \label{orbit2}
\end{equation}%
Now, imposing the equivalence between these last two equations we obtain
\begin{equation}
\frac{\partial F}{\partial \phi }=H(\phi )W_{\chi }\quad ,\quad \frac{%
\partial F}{\partial \chi }=-H(\phi )W_{\phi }.  \label{partialF}
\end{equation}%
It is important to observe that, the above equation provides a necessary
condition to get a number of new interesting models, as we are going to see
below. Once we determine $H(\phi )$, it can be replaced in (\ref{partialF})
and, then, by direct integration we can obtain a solution for the orbit
equation. In order to ensure that $dF(\phi ,\chi )$ is an exact
differential, the following constraint must be fulfilled
\begin{equation}
\frac{\partial ^{2}F}{\partial \phi \partial \chi }=\frac{\partial ^{2}F}{%
\partial \chi \partial \phi }.
\end{equation}%
This last equation, along with (\ref{partialF}), provides a definition of $%
H(\phi )$. So, applying the above condition in (\ref{partialF}), we are led
to the following relation between the superpotential and the integrating
factor

\begin{equation}
\frac{W_{\phi \phi }+W_{\chi \chi }}{W_{\phi }}=-\frac{d\ln H(\phi )}{d\phi }%
.
\end{equation}%
It can be observed that the right-hand side of the above equation depends
only on $\phi $. Therefore, the left-hand side must be just a function of $%
\phi $ too. Thus, one might establish the following condition of
applicability of the method
\begin{equation}
\frac{W_{\phi \phi }+W_{\chi \chi }}{W_{\phi }}=f(\phi ).  \label{condition}
\end{equation}%
By using the above condition, we identify a class of superpotentials that
could be studied with this formalism. In fact, a large amount of models
already considered in the literature satisfies the condition (\ref{condition}%
) \cite{bnrt,Dutra1,bazeiabrito,prieslei}. Moreover, the last equation could be
integrated to give the integrating factor
\begin{equation}
H(\phi )=e^{-\int d\phi f(\phi )}.  \label{integrating}
\end{equation}

At this point, it is important to stress that the equation (\ref{condition})
establishes a condition which is necessary and sufficient to construct the
orbit equations of the new models that we are going to present in the next.
However, it is also important to note that it is still necessary\ to write
one field as a function of the other one, in order to obtain analytical
solutions, and this restricts the\ set of useful orbits.

In order to exemplify the procedure above described, let us consider the so
called BNRT model \cite{bnrt}. In this case, the superpotential is given by
\begin{equation}
W(\phi ,\chi )=\lambda \bigg(\frac{1}{3}\chi ^{3}-a^{2}\chi \bigg)+\mu \chi
\phi ^{2},
\end{equation}%
Now, by checking that the condition (\ref{condition}) is satisfied, one gets
\begin{equation}
\frac{W_{\phi \phi }+W_{\chi \chi }}{W_{\phi }}=\frac{1+\lambda /\mu }{\phi }%
=f(\phi ).
\end{equation}%
Then, the integrating factor can be determined by using equation (\ref%
{integrating})
\begin{equation}
H(\phi )=\exp {\bigg(-\int d\phi \frac{1+\lambda /\mu }{\phi }\bigg)}=\phi
^{-(1+\lambda /\mu )}
\end{equation}%
by inserting this result in (\ref{partialF}) we arrive at
\begin{equation}
\frac{\partial F}{\partial \phi }=\frac{\lambda (\chi ^{2}-a^{2})+\mu \phi
^{2}}{\phi ^{(1+\lambda /\mu )}}\quad ,\quad \frac{\partial F}{\partial \chi
}=-2\mu \chi \phi ^{-\lambda /\mu }.
\end{equation}%
Finally, by direct integration, we obtain
\begin{equation}
F(\phi ,\chi )=\mu (\chi ^{2}-a^{2})\phi ^{-\lambda /\mu }-\frac{\mu }{%
2-\lambda /\mu }\phi ^{2-\lambda /\mu }=c,
\end{equation}%
which is the implicit form of the solution for the orbit equation. It is
interesting to note that the above solution is the same that was obtained in
ref. \cite{Dutra1,Shifman}. However, since the solutions for the above model had
already been studied in the literature we will not discuss it here.

\section{Generating new nonlinear models}

In this section we are going to present a systematic procedure that enables
us to obtain new nonlinear scalar field models that satisfies the condition (\ref{condition}) and, as a consequence, the orbit equation of such systems
arises naturally from the exact differential method. Essentially the
procedure that we are going to introduce here consists in
getting the solution of the equation 
\begin{equation}\label{condition3}
W_{\phi \phi }+W_{\chi \chi } = f(\phi)W_{\phi }.
\end{equation}
It is important to stress out that we are not interest in obtain a general solution for the above partial differential equation with some boundary condition, in this paper we are interest in construct simple solutions in a systematic way engendering physically interesting model for doublets of scalar fields.

\subsection{Polynomial model I:}

Let us introduce the procedure through a concrete example. Consider the
following \textit{ansatz} for the general form of the superpotential
\begin{equation}  \label{superpotential}
W(\phi ,\chi )=a_{30}\phi ^{3}+a_{31}\phi ^{2}\chi +a_{32}\phi \chi
^{2}+a_{33}\chi ^{3}+a_{10}\phi +a_{11}\chi ,
\end{equation}%
where the coefficients are arbitrary. Note that the above superpotential do
not presents any term with fourth degree. However, the respective potential
function $V(\phi ,\chi )$ will have it. Substituting the last expression in (%
\ref{condition3}) we obtain
\begin{equation}  \label{condition2}
(2a_{31}+6a_{33})\chi +(6a_{30}+2a_{32})\phi =f(\phi )[a_{32}\chi
^{2}+2a_{31}\phi \chi +3a_{30}\phi ^{2}+a_{10}],
\end{equation}%
comparing the coefficients of the above equation with respect to the powers
of $\chi $, one can conclude that $a_{32}=0$ and $2a_{31}+6a_{33}=2~a_{31}~%
\phi ~f(\phi )$. The last equation provide us a structure for the function $%
f(\phi )$, namely
\begin{equation}
f(\phi )=\frac{1+3a_{33}/a_{31}}{\phi },
\end{equation}%
replacing it in equation (\ref{condition2}) we may obtain $2a_{30}\phi
^{2}=(3a_{30}\phi ^{2}+a_{10})(1+3a_{33}/a_{31})$. Comparing the
coefficients with respect to the powers of $\phi $, we may obtain $a_{10}=0$%
, $a_{31}=3a_{33}$, and consequently $f(\phi )=2/\phi $. Note that the other
coefficients ($a_{11},a_{30},a_{31}$) remain free. Thus, the superpotential (%
\ref{superpotential}) may be rewritten as follows
\begin{equation}
W(\phi ,\chi )=a_{30}\phi ^{3}+a_{31}\bigg(\phi ^{2}\chi +\frac{1}{3}\chi
^{3}\bigg)+a_{11}\chi .
\end{equation}%
This superpotential generalizes the so called BNRT model \cite{bnrt}, note
that we recover the BNRT case when $a_{30}=0$. The corresponding potential
is
\begin{equation*}
V(\phi ,\chi )=\frac{1}{2}[3a_{30}\phi ^{2}+2a_{31}\phi \chi ]^{2}+\frac{1}{2%
}[a_{31}(\phi ^{2}+\chi ^{2})+a_{11}]^{2}.
\end{equation*}%
Using the same approach already realized in the previous section, we may
obtain the implicit solution for the orbit equation
\begin{equation}
F(\phi ,\chi )=a_{31}\bigg(\phi -\frac{\chi ^{2}}{\phi }\bigg)-\frac{a_{11}}{%
\phi }-3a_{30}\chi =c.  \label{orbit3}
\end{equation}

Now, let us look for a solution of this model. In this case we will restrict
ourselves to the identification $a_{11}=-a_{31}=1$, and also, we identify $%
a_{30}=-\beta /3$. So, the superpotential function may be rewritten as
follows
\begin{equation}
W(\phi ,\chi )=\chi -\frac{1}{3}\chi ^{3}-\chi \phi ^{2}-\frac{\beta }{3}%
\phi ^{3},
\end{equation}%
The first order equation derived from this superpotential may be written as
\begin{equation}
\frac{d\phi }{dx}=-2\chi \phi -\beta \phi ^{2}\quad ,\quad \frac{d\chi }{dx}%
=1-\chi ^{2}-\phi ^{2}.  \label{firstorder2}
\end{equation}%
In fact, this superpotential is an asymmetric version of the so called BNRT
model discussed above, and it is not difficult to see that the addition of
the term $\beta \phi ^{3}/3$ in the superpotential, breaks the $Z_{2}\times
Z_{2}$ symmetry that is presented in the potential of the BNRT model. The
vacua of the model may be obtained, as usual, from $W_{\phi }=W_{\chi }=0$.
In this case we get four vacua corresponding to the coordinates $(\phi
_{v},\chi _{v})$ in the internal space
\begin{eqnarray}
v_{1} &=&(0,1)\qquad v_{2}=\bigg(\frac{2}{\sqrt{\beta ^{2}+4}},\frac{-\beta
}{\sqrt{\beta ^{2}+4}}\bigg)\qquad v_{3}=(0,-1)\quad  \notag \\
&& \\
v_{4} &=&\bigg(\frac{-2}{\sqrt{\beta ^{2}+4}},\frac{\beta }{\sqrt{\beta
^{2}+4}}\bigg).  \notag
\end{eqnarray}%
By using equation (\ref{orbit3}) we can express the orbit equation as
follows
\begin{equation}
\chi ^{2}-1+\beta \phi \chi =\phi ^{2}-b\sqrt{\beta ^{2}+4}\phi ,
\label{orbit4}
\end{equation}
where the identification $c\equiv -b\sqrt{\beta ^{2}+4}$ was used. It is
interesting to note that the vacua states $v_{1}$ and $v_{3}$ always satisfy
the above equation, independently of the values of $b$. However, the other
two vacua, namely $v_{2}$ and $v_{4}$, only satisfy (\ref{orbit4}) if the
parameter $b$ is taken to be equal to the following critical values $b=\pm 1$.

\begin{figure}[H]
\label{figorbit1} \centering
\includegraphics[scale=.8]{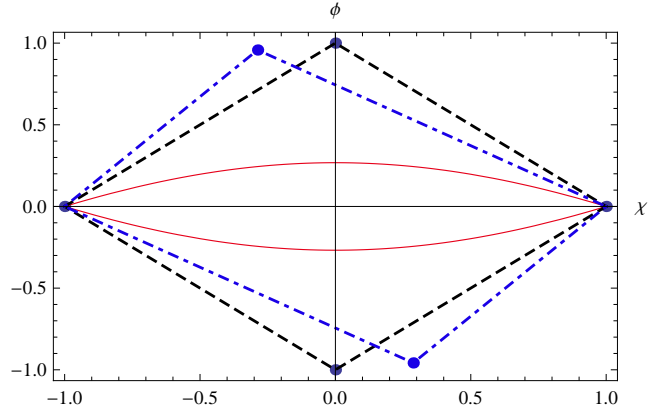}
\caption{Vacua structure and orbits solutions. $\protect\beta = 0$ and $b =
2 $ - solid line (red); $\protect\beta = 0$ and $b = 1,000000001$ - dashed
line (black); $\protect\beta = 0,6$ and $b = 1,000000001$ - dotdashed line
(blue);}
\end{figure}

\indent In order to decouple the pair of first order equations (\ref{firstorder2})
one can use the orbit (\ref{orbit4}) to express $\phi $ as a function of $%
\chi $, so that
\begin{equation}
\phi (\chi )=\frac{\beta \chi +b\sqrt{\beta ^{2}+4}-f(\chi )}{2},
\label{orbit5}
\end{equation}%
where $f(\chi )=\sqrt{(\chi \sqrt{\beta ^{2}+4}+\beta b)^{2}+4(b^{2}-1)}$.
Substituting it in the second equation of (\ref{firstorder2}) and performing
the integration, we obtain the following solution
\begin{equation*}
\chi (x)=\sqrt{\frac{b^{2}-1}{\beta ^{2}+4}}\bigg[\frac{2\tanh (x-x_{0})+2b}{%
\sqrt{b^{2}-1}(\sqrt{\beta ^{2}+4}-\beta )}-\frac{\sqrt{b^{2}-1}(\sqrt{\beta
^{2}+4}-\beta )}{2\tanh (x-x_{0})+2b}-\frac{\beta b}{\sqrt{b^{2}-1}}\bigg].
\end{equation*}%
The other field, $\phi (x)$, can be obtained by direct substitution of the
explicit form of $\chi (x)$ in the equation (\ref{orbit5}). As one can see
in Figure 2, the field $\chi (x)$ presents a kink-like behavior while $\phi
(x)$, exhibits a lump-like profile. It is a remarkable fact that when $b=\pm
(1+\varepsilon )$ (with $\varepsilon $ being a positive and very small
parameter), the field $\chi (x)$ develops a two-kink behavior, and $\phi (x)$%
, shows a flat-top region on the lump structure. In the exact situation with
$b=\pm 1$, both fields present a kink-like structure. It is interesting to
note that one can recover the results obtained in \cite{Dutra2} by choosing $%
\beta =0$.

\begin{figure}[H]
\label{kink1} \centering
\includegraphics[scale=1.4]{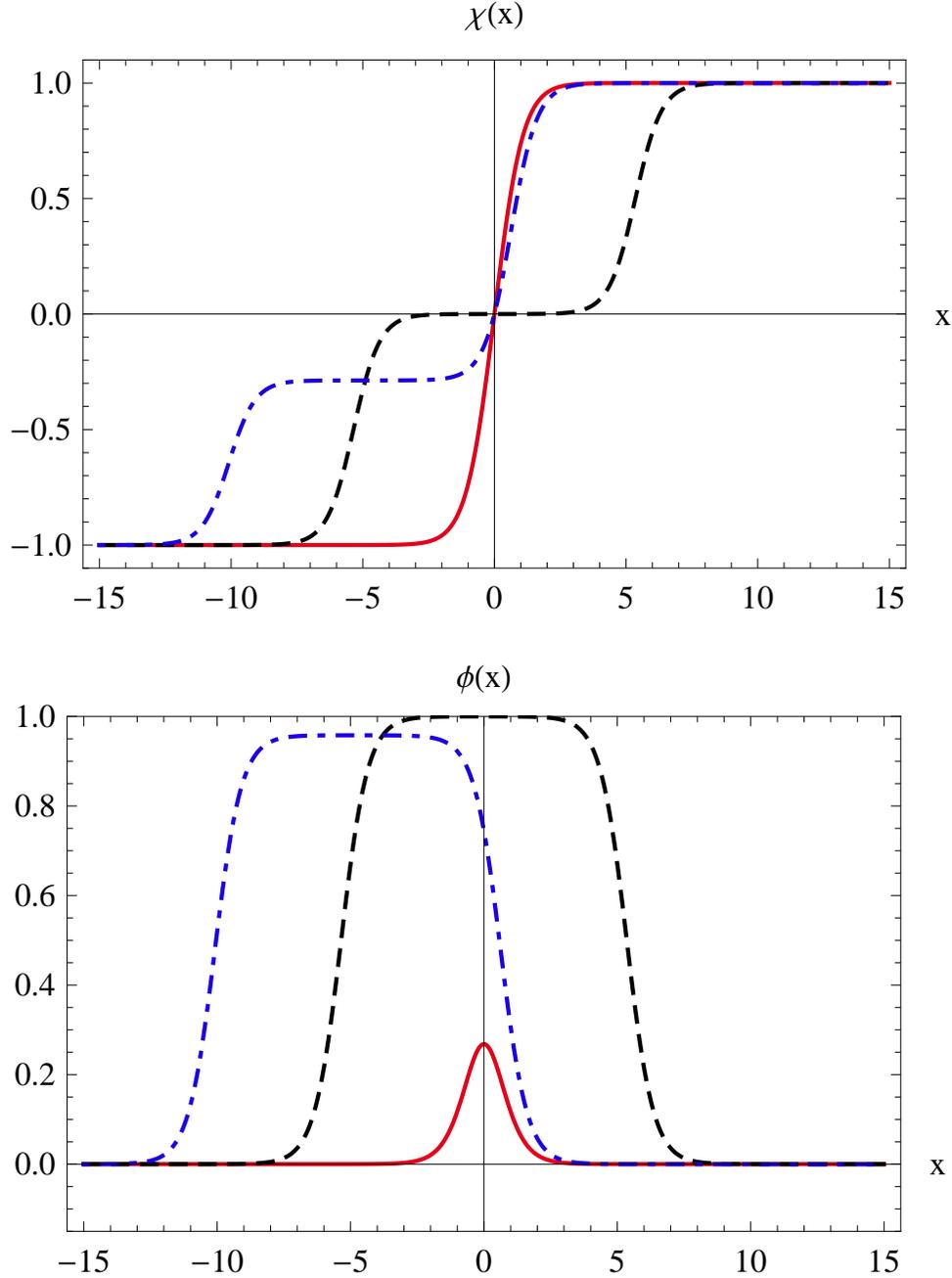}
\caption{Double-kink (upper panel) and a flat-top lump. $\protect\beta = 0$
and $b = 2$ - solid line (red); $\protect\beta = 0$ and $b = 1,000000001$ -
dashed line (black); $\protect\beta = 0,6$ and $b = 1,000000001$ - dotdashed
line (blue);}
\end{figure}

\subsection{Polynomial model II:}

The second model to be considered here is a polynomial superpotential with
fourth power degree terms (the corresponding potential $V(\phi ,\chi )$ will
contain terms with power of sixth degree in the fields). For this, we
consider the following structure for the superpotential
\begin{equation*}
W(\phi ,\chi )=a_{40}\phi ^{4}+a_{41}\phi ^{3}\chi +a_{42}\phi ^{2}\chi
^{2}+a_{43}\phi \chi ^{3}+a_{44}\chi ^{4}+a_{20}\phi ^{2}+a_{21}\phi \chi
+a_{22}\chi ^{2}.
\end{equation*}%
Repeating the same procedure used in the previous section we may conclude
that $f(\phi )=2/\phi $ and consequently $H(\phi )=1/\phi ^{2}$. Adjusting
the coefficients by the same method of the previous section and then
substituting it in the superpotential, we get
\begin{equation}
W(\phi ,\chi )=a_{40}\bigg(\phi ^{4}-2\phi ^{2}\chi ^{2}-\frac{1}{3}\chi ^{4}%
\bigg)+a_{41}\phi ^{3}\chi +a_{20}(\phi ^{2}+\chi ^{2}),
\end{equation}
and we get the following potential
\begin{equation*}
V(\phi ,\chi )=\frac{1}{2}[a_{40}(4\phi ^{3}-4\phi \chi ^{2})+3a_{41}\phi
^{2}\chi +2a_{20}\phi ]^{2}+\frac{1}{2}[a_{40}(-4\phi ^{2}\chi -4\chi
^{3}/3)+a_{41}\phi ^{3}+2a_{20}\chi ]^{2}.
\end{equation*}

By replacing the expressions above obtained in (\ref{partialF}), and
integrating them in their respective variables, we may obtain that the
implicit solution for the orbit equation that is given by
\begin{equation}  \label{orbit6}
F(\phi ,\chi )=a_{40}\bigg(\frac{4\chi ^{3}}{3\phi }-4\phi \chi \bigg)+\frac{%
a_{41}}{2}\phi ^{2}-2a_{20}\frac{\chi }{\phi }-\frac{3a_{41}}{2}\chi ^{2}=c.
\end{equation}

Now, we are going to look for solutions of this model. We will restrict
ourselves to the case where $a_{41}=0$, and also, us identify $a_{40}=1/4$
and $a_{20}=\beta ^{2}/6$. Then the superpotential function can be rewritten
as follow
\begin{equation*}
W(\phi ,\chi )=\frac{1}{4}\bigg(\phi ^{4}-2\phi ^{2}\chi ^{2}-\frac{1}{3}%
\chi ^{4}\bigg)+\frac{\beta ^{2}}{6}(\phi ^{2}+\chi ^{2}).
\end{equation*}%
The corresponding first order differential equations are given by
\begin{equation}
\frac{d\phi }{dx}=\phi ^{3}-\phi \chi ^{2}+\frac{\beta ^{2}}{3}\phi \quad
,\quad \frac{d\chi }{dx}=-\phi ^{2}\chi -\frac{1}{3}\chi ^{3}+\frac{\beta
^{2}}{3}\chi ,  \label{firstorder3}
\end{equation}%
and the orbit (\ref{orbit6}) turns out to be
\begin{equation}
\left( \frac{\beta ^{2}+3\phi ^{2}-\chi ^{2}}{3}\right) \chi -\frac{\beta
^{2}b}{\sqrt{3}}\phi =0,
\end{equation}%
where the redefinition $c\equiv -\beta ^{2}b/\sqrt{3}$ was used. The
potential $V(\phi ,\chi )$ possess seven different vacua states that are
given by
\begin{eqnarray}
v_{1} &=&(0,0)\quad v_{2}=\bigg(0,\sqrt{\beta }\bigg)\quad v_{3}=\bigg(0,-%
\sqrt{\beta }\bigg)\quad v_{4}=\bigg(\frac{\beta }{\sqrt{6}},\frac{\beta }{%
\sqrt{2}}\bigg)  \notag \\
v_{5} &=&\bigg(-\frac{\beta }{\sqrt{6}},-\frac{\beta }{\sqrt{2}}\bigg)\quad
v_{6}=\bigg(-\frac{\beta }{\sqrt{6}},\frac{\beta }{\sqrt{2}}\bigg)\quad
v_{7}=\bigg(\frac{\beta }{\sqrt{6}},-\frac{\beta }{\sqrt{2}}\bigg).
\end{eqnarray}%
In the Figure 3 we plot the vacua structure and some possible orbits.

\begin{figure}[H]
\centering
\includegraphics[scale=.8]{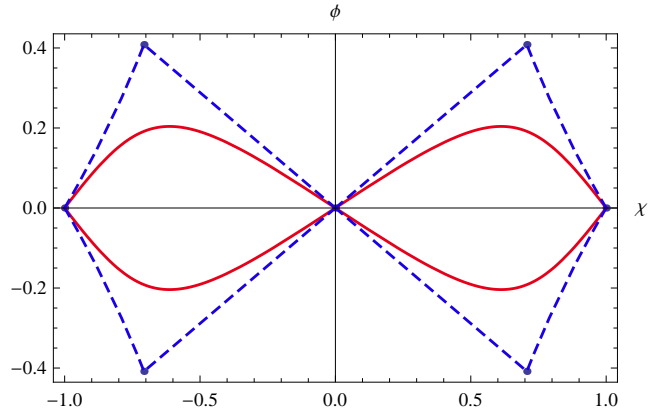}
\caption{Vacua structure and orbits solutions. $\protect\beta = 1$ and $b =
1,3$ - solid line (red); $\protect\beta = 1$ and $b = 1,00000001$ - dashed
line (blue);}
\label{fig:digraph}
\end{figure}

\indent An interesting fact is that it is possible to obtain analytical
solutions for this model, unlike other models with sixth degree terms on its
potential \cite{meza}. Note that it is possible to use the above orbit in
order to express $\phi $ in terms of $\chi $ and then substitute it in (\ref%
{firstorder3}). Performing some changes of variables we may integrate the
remaining first order equation to obtain
\begin{equation}
\chi (x)=\pm \frac{\beta }{2}\bigg[2+b+\tanh (\beta ^{2}(x-x_{0})/3)-\frac{%
b^{2}-1}{\tanh (\beta ^{2}(x-x_{0})/3)+b}\bigg]^{1/2},
\end{equation}
the field $\phi (x)$ can be determined by direct substitution of the last
equation into the orbit equation, resulting into the following expression
\begin{equation}
\phi (\chi )=\frac{b\beta ^{2}}{2\sqrt{3}\chi }\bigg[1-\sqrt{1+\frac{4\chi
^{2}}{b^{2}\beta ^{4}}(\chi ^{2}-\beta ^{2})}\bigg]
\end{equation}%
In the Figure 4 we plot the solutions mentioned above for some specific
values of $\beta $ and $b$, note that the field $\chi (x)$ present an
asymmetrical two-kink like profile when the integration constant is close to
a certain critical value, while the other field $\phi (x)$ exhibits a
lump-like solution with a flat top region.

\begin{figure}[H]
\centering
\includegraphics[scale=1.4]{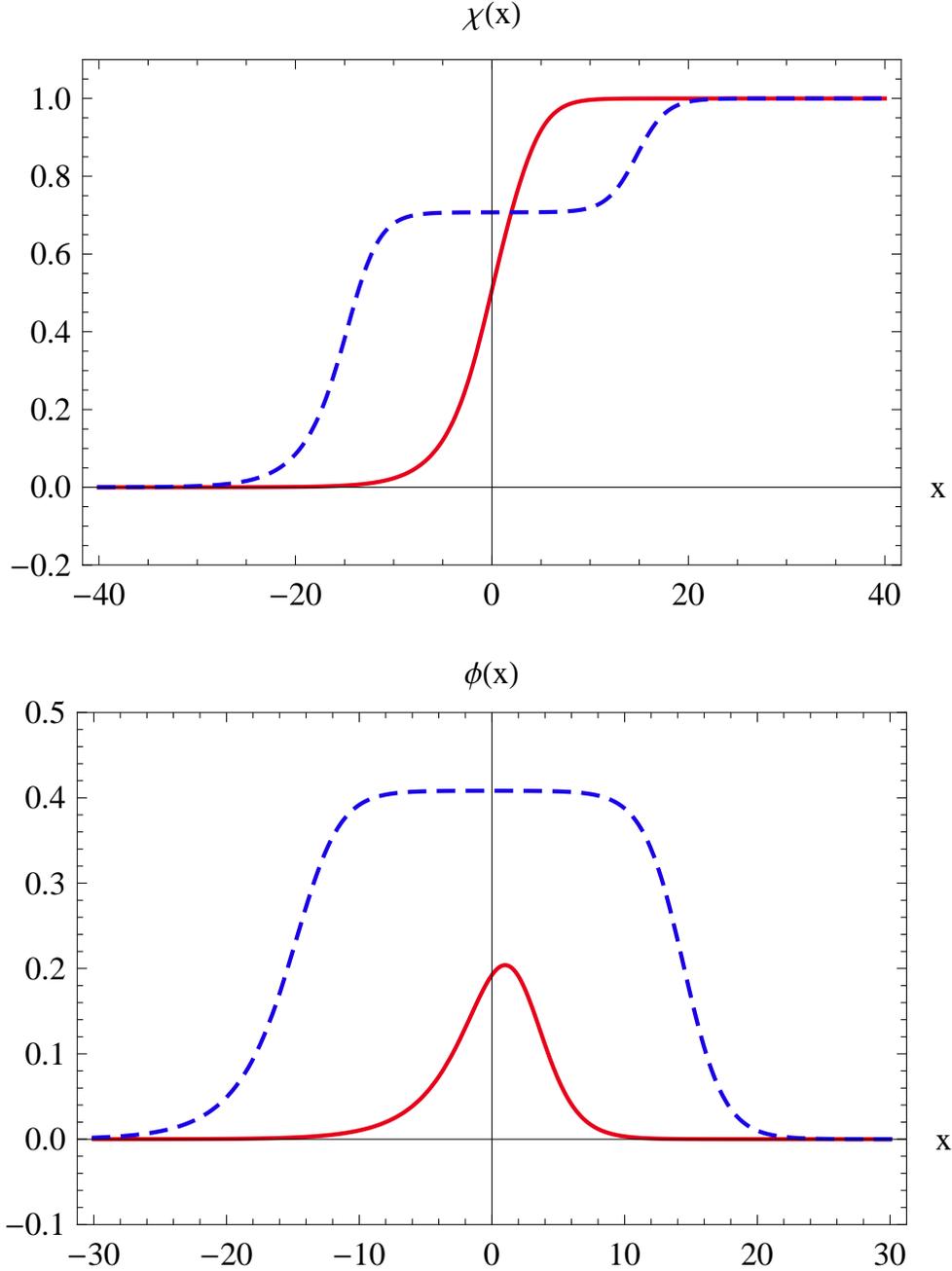}
\caption{Fields solutions for the polynomial model II. $\protect\beta = 1$
and $b = 1,3$ - solid line (red); $\protect\beta = 1$ and $b = 1,00000001$ -
dashed line (blue);}
\label{fig:digraph2}
\end{figure}

\subsection{Polynomial model III:}

\indent The third model we consider is characterized by a polynomial
superpotential containing terms with fifth power degree, whose corresponding
potential $V(\phi ,\chi )$ contains eighth degree terms. Let us consider the
following structure for the superpotential
\begin{eqnarray}
W(\phi ,\chi ) &=&a_{50}\phi ^{5}+a_{51}\phi ^{4}\chi +a_{52}\phi ^{3}\chi
^{2}+a_{53}\phi ^{2}\chi ^{3}+a_{54}\phi \chi ^{4}+a_{55}\chi ^{5}+  \notag
\\
&&+a_{30}\phi ^{3}+a_{31}\phi ^{2}\chi +a_{32}\phi \chi ^{2}+a_{33}\chi
^{3}+a_{10}\phi +a_{11}\chi .
\end{eqnarray}%
Using the same procedure of the previous sections we obtain, once more, that
$f(\phi )=2/\phi $ and, as a consequence, $H(\phi )=1/\phi ^{2}$. Adjusting
the coefficients by the same method of the previous section and then
substituting it in the superpotential, one may conclude that
\begin{equation*}
W(\phi ,\chi )=a_{50}(\phi ^{5}-5\phi ^{3}\chi ^{2})+a_{31}\bigg(\phi
^{2}\chi +\frac{1}{3}\chi ^{3}\bigg)+a_{30}\phi ^{3}+a_{51}\bigg(\phi
^{4}\chi -\frac{2}{3}\phi ^{2}\chi ^{3}-\frac{1}{15}\chi ^{5}\bigg)%
+a_{11}\chi .
\end{equation*}%
The implicit solution for the orbit equation is
\begin{equation*}
F(\phi ,\chi )=5a_{50}(\chi ^{3}-\phi ^{2}\chi )+a_{51}\bigg(\frac{1}{3}\phi
^{3}-2\phi \chi ^{2}+\frac{\chi ^{4}}{3\phi }\bigg)+a_{31}\bigg(\phi -\frac{%
\chi ^{2}}{\phi }\bigg)-3a_{30}\chi -\frac{a_{11}}{\phi }=c.\quad
\end{equation*}%
\indent Let us study the particular case with $a_{50}=a_{30}=0$, $a_{51}=1$,
$a_{31}=\gamma $ and $a_{11}=\beta /4$ (with $\gamma $ and $\beta $
positives). Unfortunately, in this case, it will be possible to carry out
analytical calculations only partially. However, it is interesting to
analyze this model since some interesting features will arise. Also, this
case is a good example to show the importance of obtaining an analytical
expression for the orbit solution. Substituting the specific values of the
coefficients in the superpotential function we get
\begin{equation}
W(\phi ,\chi )=\phi ^{4}\chi -\frac{2}{3}\phi ^{2}\chi ^{3}-\frac{\chi ^{5}}{%
15}+\gamma \bigg(\phi ^{2}\chi +\frac{1}{3}\chi ^{3}\bigg)+\frac{\beta }{4}%
\chi .
\end{equation}%
The corresponding first order differential equations are given by
\begin{equation}
\frac{d\phi }{dx}=4\phi ^{3}\chi -\frac{4}{3}\phi \chi ^{3}+2\gamma \phi
\chi \quad ,\quad \frac{d\chi }{dx}=\phi ^{4}-2\phi ^{2}\chi ^{2}-\frac{1}{3}%
\chi ^{4}+\gamma (\phi ^{2}+\chi ^{2})+\frac{\beta }{4},  \label{firstorder4}
\end{equation}%
and the orbit solution is
\begin{equation}
\frac{1}{3}(\phi ^{4}-6\phi ^{2}\chi ^{2}+\chi ^{4})+\gamma (\phi ^{2}-\chi
^{2})-\frac{\beta }{4}=c\phi  \label{orbit7}
\end{equation}%
\indent The potential function $V(\phi ,\chi )$ possess six vacua. In order
to specify this vacua let us define the following quantities
\begin{eqnarray}
\phi _{vac}^{(1)} &=&0\quad \phi _{vac}^{(2)}=\bigg[-\frac{\gamma }{8}+\frac{%
\sqrt{2\beta +7\gamma ^{2}}}{2}\bigg]^{1/2}\quad \chi _{vac}^{(1)}=\bigg[%
\frac{3\gamma +\sqrt{3\beta +9\gamma ^{2}}}{2}\bigg]^{1/2}\quad  \notag \\
&& \\
\chi _{vac}^{(2)} &=&\bigg[\frac{9\gamma +3\sqrt{2\beta +7\gamma ^{2}}}{8}%
\bigg]^{1/2}  \notag
\end{eqnarray}%
thus, the corresponding coordinates of the vacua states in the internal
space may be written as follows
\begin{eqnarray}
v_{1} &=&(\phi _{vac}^{(1)},\chi _{vac}^{(1)})\quad v_{2}=(\phi
_{vac}^{(1)},-\chi _{vac}^{(1)})\quad v_{3}=(\phi _{vac}^{(2)},\chi
_{vac}^{(2)})  \notag \\
v_{4} &=&(\phi _{vac}^{(2)},-\chi _{vac}^{(2)})\quad v_{5}=(-\phi
_{vac}^{(2)},\chi _{vac}^{(2)})\quad v_{6}=(-\phi _{vac}^{(2)},-\chi
_{vac}^{(2)}).
\end{eqnarray}

\begin{figure}[H]
\centering
\includegraphics[scale=.8]{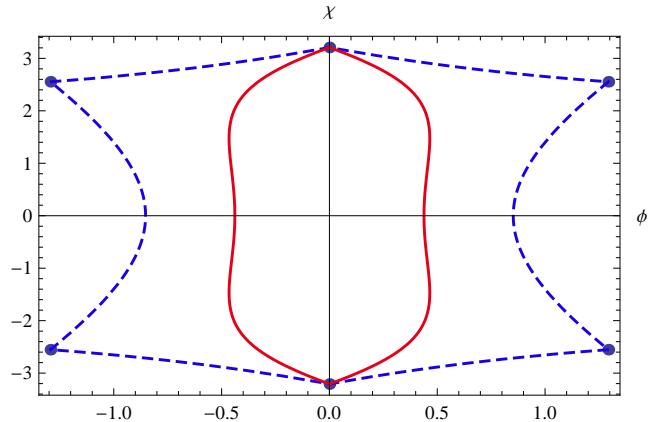}
\caption{Vacua structure and orbits solutions. $\protect\beta = 100$, $%
\protect\gamma = 1$ and $b = 2$ - solid line (red); $\protect\beta = 100$, $%
\protect\gamma = 1$ and $b = 1 + 10^{-40}$ - dashed line (blue);}
\label{fig:digraph3}
\end{figure}

\indent It is not difficult to verify that the vacua states $v_{1}$ and $%
v_{2}$ satisfy (\ref{orbit7}) independently of the value chosen for the
constant $c$. Otherwise, other vacua states satisfies the orbit (\ref{orbit7}%
) only for some critical value of $c$ that is determined by the substitution
of coordinates of the vacua sates in the orbit solution \cite{meza}. For
instance, let us consider the vacua $v_{3}$, the critical value $c_{0}$ is
given by
\begin{equation}
c_{0}=\frac{1}{\phi _{vac}^{(2)}}\bigg[\frac{1}{3}\bigg((\phi
_{vac}^{(2)})^{4}-6(\phi _{vac}^{(2)})^{2}(\chi _{vac}^{(2)})^{2}+(\chi
_{vac}^{(2)})^{4}\bigg)+\gamma \bigg((\phi _{vac}^{(2)})^{2}-(\chi
_{vac}^{(2)})^{2}\bigg)-\frac{\beta }{4}\bigg].
\end{equation}%
It is easy to see that the vacuum $v_{4}$ possess the same value for
critical constant while $v_{5}$ and $v_{6}$ possess the critical value $%
-c_{0}$. It is interesting rewrite the orbit solution in terms of this
critical parameter, namely
\begin{equation}
\frac{1}{3}(\phi ^{4}-6\phi ^{2}\chi ^{2}+\chi ^{4})+\gamma (\phi ^{2}-\chi
^{2})-\frac{\beta }{4}=b~c_{0}~\phi .  \label{orbit8}
\end{equation}%
While $c$ defines a family of orbits in equation (\ref{orbit7}), $b$ defines
a family of orbits in the above equation. In the Figure 5 we plot the orbit
solution for some values of the parameters $b$, $\gamma $ and $\beta $.
Finally, in order to perform the numerical integration in the first order
equation (\ref{firstorder4}) we have to specify initial values for both
fields, and the orbit solution is very useful at this point. For instance,
let us look for solutions that connect $v_{1}$ and $v_{2}$, certainly there
exists a point $x_{0}$ in which $\chi (x_{0})=0$ (we will choose $x_{0}=0$
without loosing generality, since the problem possess a translational
invariance). The corresponding value of $\phi $ at $x=0$ can be directly
determined from the orbit solution, by solving the following equation
\begin{equation}
\frac{1}{3}\phi (0)^{4}+\gamma \phi (0)^{2}-bc\phi (0)-\frac{\beta }{4}=0.
\end{equation}%
In the Figure 6 we plot the numerical solution obtained through this
procedure for some values of the parameters $b$, $\gamma $ and $\beta $.
Note that, for some values of $b$, the field $\chi (x)$ presents a kind of
triple kink configuration while the field $\phi (x)$ present double lump
behavior with a flat-top region. In the Figure 7 one can see that there
exists three regions with a formation of peaks in the energy density. As far
we know this kind of configuration was never presented in the literature. In
fact, very recently a solution like those was obtained in a model with one
self-interacting scalar field \cite{prd2014}. However, beyond the fact that
there is only one field in the model, the potential is not entirely
continuous, instead, it is continuous by parts. Here, the model is
absolutely continuous and the model is for a doublet of scalar fields.

\begin{figure}[H]
\centering
\includegraphics[scale=1.4]{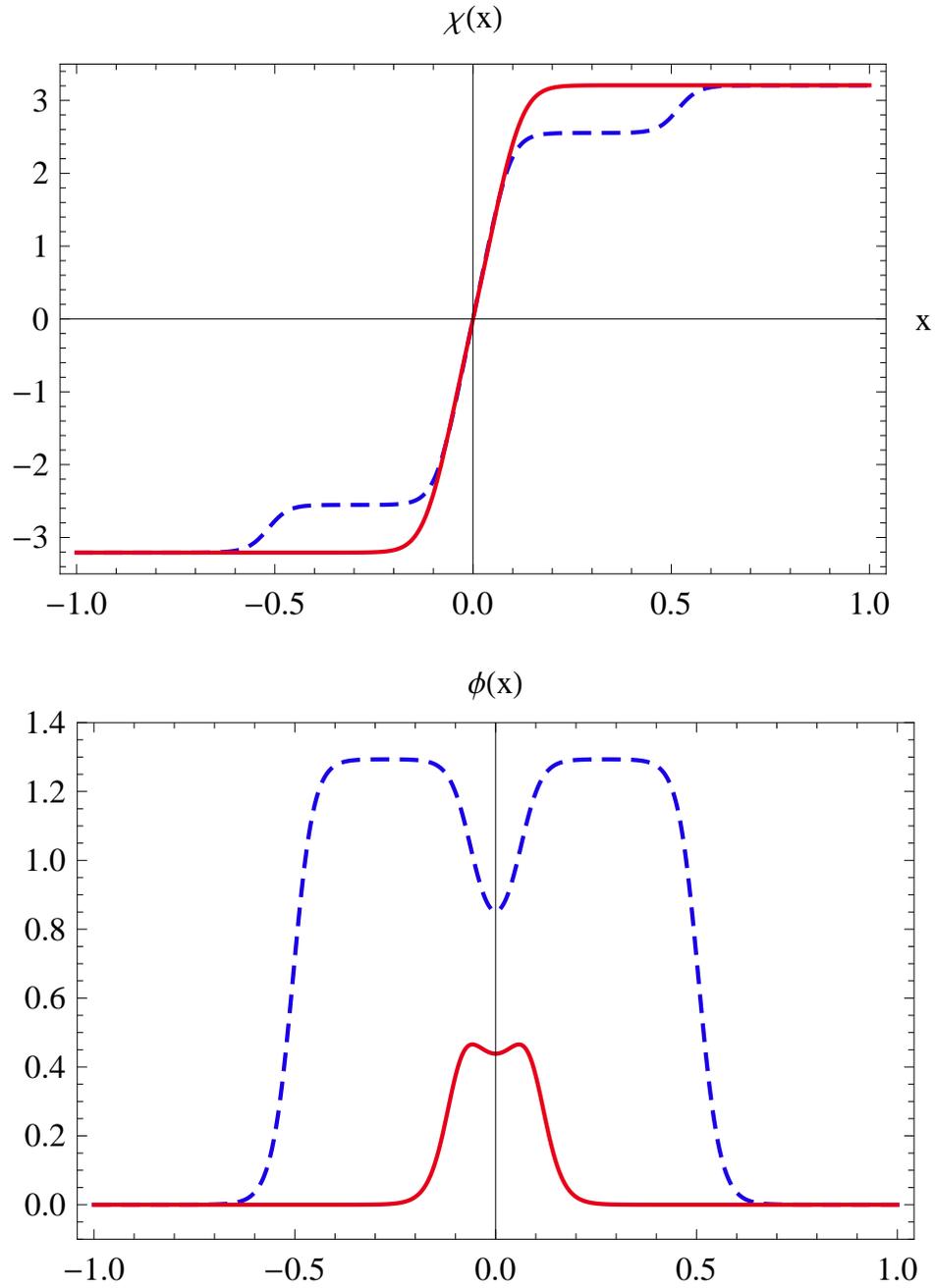}
\caption{Fields solutions for the polynomial model III. $\protect\beta = 100$%
, $\protect\gamma = 1$ and $b = 2$ - solid line (red); $\protect\beta = 100$%
, $\protect\gamma = 1$ and $b = 1 + 10^{-40}$ - dashed line (blue);}
\label{fig:digraph4}
\end{figure}

\begin{figure}[H]
\centering
\includegraphics[scale=.8]{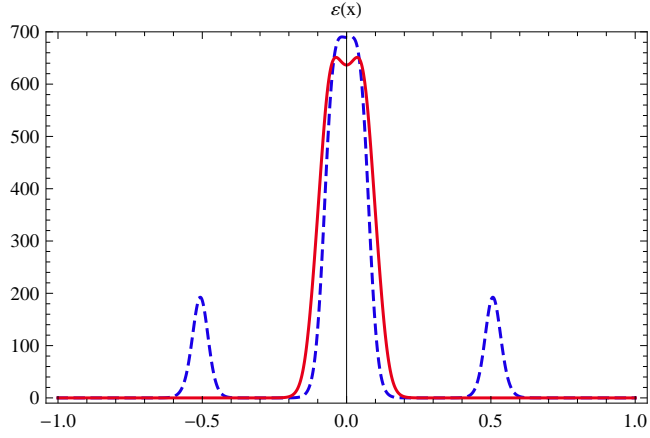}
\caption{Energy density for the polynomial model III. $\protect\beta = 100$,
$\protect\gamma = 1$ and $b = 2$ - solid line (red); $\protect\beta = 100$, $%
\protect\gamma = 1$ and $b = 1 + 10^{-40}$ - dashed line (blue);}
\label{fig:digraph5}
\end{figure}

\subsection{Generalized polynomial model:}

In this section we are going to generalize the procedure exemplified in the
preceding sections and obtain a polynomial superpotential with Nth degree
terms. For this we consider the following superpotential
\begin{equation}
W^{(N)}(\phi ,\chi )=\sum_{n=0}^{N}W^{(n)}(\phi ,\chi ),
\end{equation}%
where
\begin{equation}
W^{(n)}(\phi ,\chi )=\sum_{l=0}^{n}a_{nl}\phi ^{n-l}\chi ^{l}.  \label{Wn}
\end{equation}

Note that the equation $W_{\phi \phi }^{(N)}+W_{\chi \chi }^{(N)}=f(\phi
)W_{\phi }^{(N)}$ is linear in $W^{(N)}$. Therefore, one can solve the
equations $W_{\phi \phi }^{(n)}+W_{\chi \chi }^{(n)}=f(\phi )W_{\phi }^{(n)}$
individually and, then, sum over its solutions in order to obtain $W^{(N)}$.
The three models considered previously in this work provide us the result $%
f(\phi )=2/\phi $. Thus, it seems interesting to consider this result in
this generalization. From now on, our task reduce to solve the following
equation
\begin{equation}
W_{\phi \phi }^{(n)}+W_{\chi \chi }^{(n)}=\frac{2}{\phi }W_{\phi }^{(n)}.
\end{equation}

Substituting the superpotential (\ref{Wn}) in the above equation, and
repeating the same procedure of the previous sections, in other words,
comparing the coefficients accordingly to the degree of $\phi $ and $\chi $,
we may obtain $a_{n(n-1)}=0$ and also the following recurrence formula
\begin{equation}
a_{n(l+2)}=\frac{(n-l)(3+l-n)}{(l+2)(l+1)}a_{nl}.  \label{recursion}
\end{equation}%
by successive applications of the above recurrence relation, we may find
that the general term is given by
\begin{equation}
a_{nl}=\frac{n!!}{l!(n-l)!!}\prod_{k=0}^{(l-2)/2}(3+2k-n)a_{n0},
\end{equation}%
for $l$ even, and
\begin{equation}
a_{nl}=\frac{(n-1)!!}{l!(n-l)!!}\prod_{k=0}^{(l-3)/2}(4+2k-n)a_{n1},
\end{equation}%
for $l$ odd. Above we have used $n!!=n(n-2)!!$.

Now we turn our attention to the solution of the orbit equation, which is in
general nonlinear in terms of the fields. However, it is linear in terms of
the implicit solution $F(\phi ,\chi )$. Therefore, we may look for solutions
in the form
\begin{equation}
F(\phi ,\chi )=\sum_{n=0}^{N}F^{(n)}(\phi ,\chi )=c,
\end{equation}%
where the function $F^{(n)}(\phi ,\chi )$ satisfies the following equation
\begin{equation*}
dF^{(n)}(\phi ,\chi )=\frac{\partial F^{(n)}}{\partial \phi }d\phi +\frac{%
\partial F^{(n)}}{\partial \chi }d\chi =H(\phi )W_{\chi }^{(n)}d\phi -H(\phi
)W_{\phi }^{(n)}d\chi =0.
\end{equation*}%
By comparison of the terms in the above equation, we find that
\begin{equation}
\frac{\partial F^{(n)}}{\partial \phi }=H(\phi )W_{\chi }^{(n)},~\frac{%
\partial F^{(n)}}{\partial \chi }=-H(\phi )W_{\phi }^{(n)}.
\end{equation}%
Integrating the above equations and taking into account the recurrence
relation (\ref{recursion}), we may obtain the following result
\begin{equation}
F^{(n)}(\phi ,\chi )=\sum_{l=1}^{n}\frac{l}{n-l-1}a_{nl}\phi ^{n-l-1}\chi
^{l-1}+\beta ^{(n)}(\chi )=c_{n},
\end{equation}%
where
\begin{equation}
\beta ^{(n)}(\chi )=%
\begin{cases}
0 & ,n<3 \\
\frac{3}{2-n}a_{n(n-3)}\chi ^{n-2} & ,n\geq 3.%
\end{cases}
\label{potecial_2}
\end{equation}%
summing over all the possible values of $n$, we get
\begin{equation*}
F(\phi ,\chi )=\sum_{n=1}^{N}\sum_{l=1}^{n}\frac{l}{n-l-1}a_{nl}\phi
^{n-l-1}\chi ^{l-1}+\sum_{n=1}^{N}\beta ^{(n)}(\chi )=c.
\end{equation*}%
Note that
\begin{equation}
\sum_{n=1}^{N}\beta ^{(n)}=\sum_{n=3}^{N}\frac{3}{2-n}a_{n(n-3)}\chi ^{n-2},
\end{equation}%
thus, we obtain
\begin{equation*}
F(\phi ,\chi )=\sum_{n=1}^{N}\sum_{l=1}^{n}\frac{l}{n-l-1}a_{nl}\phi
^{n-l-1}\chi ^{l-1}+\sum_{n=3}^{N}\frac{3}{2-n}a_{n(n-3)}\chi ^{n-2}=c.
\end{equation*}

Naturally, we cannot obtain an analytical solution for the above model.
However, as it was pointed out in the previous section, the knowledge of an
analytical expression for the orbit is an important step for the analysis of
nonlinear scalar field theories. This happens because it allows one to
decouple the first order differential equations and choosing adequately
boundary conditions.

\subsection{Nonlinear oscillating models}

The systematic procedure developed in the last sections with polynomial
models, can be extended to build up models with potentials presenting
harmonic functions of the fields. For instance, let us consider the
following \textit{ansatz} for an oscillating superpotential
\begin{equation*}
W(\phi ,\chi )=A\sin \phi \sin \chi +B\cos \phi \cos \chi +C\sin \phi \cos
\chi +D\cos \phi \sin \chi +E\phi +F\chi .
\end{equation*}%
Following the same procedure of the previous sections, one can substitute
the above superpotential in (\ref{condition3}), and by comparing the involved
terms, one may obtains that
\begin{equation}
f(\phi )=\frac{2(A\sin \phi +D\cos \phi )}{D\sin \phi -A\cos \phi },
\end{equation}%
where the coefficients must obey the following constraints: $AB=CD$, $E=0$.
In the case where $C\neq 0$ we get $D=AB/C$. Thus, the superpotential can be
rewritten as
\begin{equation}
W(\phi ,\chi )=A(\sin \chi +C\cos \chi )(\sin \phi +B/C\cos \phi )+F\chi .
\end{equation}%
The integrating factor obtained by using $f(\phi )$ is given by
\begin{equation}
H(\phi )=\frac{1}{A^{2}(B/C\sin \phi -\cos \phi )^{2}},
\end{equation}%
consequently, we get the following orbit
\begin{equation*}
F(\phi ,\chi )=\frac{C\sin \chi -A\cos \chi }{A^{2}(B/C\sin \phi -\cos \phi )%
}+\frac{F/A^{2}}{B/C(1-B/C\tan ^{2}\phi )}=c.
\end{equation*}%
As far as we know, this superpotential was not considered in the literature.
However, one can note that in the case where $C=0$ we may recover the
oscillating model studied in ref. \cite{prieslei}.

\section{Conclusions}

In this work, we introduced a method which allows one obtain the
solutions of the orbit equation for the case of nonlinearly coupled two
scalar fields and, beyond that, we present a procedure that allows the
construction of new exact nonlinear models of this nature systematically, in
such a way that the solution of the orbit equation appears naturally. By
applying the method we have studied, some novel polynomial models were
introduced and we explored the behavior of their solitonic configurations.
We have also verified that this procedure can be extended to the case of
oscillating potentials.

In the first model analyzed, we got a generalization of the BNRT \cite{bnrt} model, which presents a structure with four vacua and a parameter
which controls the asymmetry of the position of those vacua. It is
noteworthy that this model presents important consequences in the braneworld
scenario \cite{hoff}. The second model which was analyzed here, was one of
the sixth degree in the fields and in analytically solvable in contrast with
happens with some others in the literature \cite{meza}. It is remarkable that the third
polynomial potential introduced in this work, exemplifying of the powerness
of the approach, presents configurations with a triple kink and a kind of
double-flat-top lump.

\bigskip

\textbf{Acknowledgements: }The authors thanks to CNPq and FAPESP for partial financial support.

\end{document}